\documentstyle[12pt]{article}
\normalbaselines
\begin{document}
\pagestyle{empty}

\vbox {\vspace{6mm}}

\begin{center}
{\large \bf COHERENT STATES FOR KRONECKER PRODUCTS OF NON COMPACT GROUPS: 
\\[2mm]
FORMULATION AND APPLICATIONS}\\[7mm]
Bindu A. Bambah\\                                                         
{\it School of Physics\\ University Of Hyderabad
Hyderabad-500134,India}\\[5mm]
G.S. Agarwal\\
{\it Physical Research Labratory , Ahmedabad}\\[5mm]
\end{center}

\vspace{2mm}

\begin{abstract}
We introduce and study the properties of a class of coherent states for
the group SU(1,1) X  SU(1,1)  and derive explicit expressions for these using 
the Clebsch-Gordan algebra for the SU(1,1) group. We restrict ourselves to the 
discrete series representations of SU(1,1). These are the generalization of the 
`Barut Girardello' coherent states to the Kronecker Product of two non-compact 
groups.The resolution of the identity and the analytic phase space 
representation of these states is presented. This  phase space representation 
is based on the basis of products of `pair coherent states' rather than the 
standard number state canonical basis. We discuss the  utility of the resulting 
`bi-pair coherent states' in  the context of four-mode interactions in quantum 
optics.
\end{abstract}
\section{FORMULATION}
\subsection{Coupling of Pair coherent states in the fock state basis}
For two mode systems the traditional SU(1,1) coherent states which have been 
extensively studied in the context of squeezing have been
 the Caves-Schumaker states \cite{caves}, defined by the relation
\begin{equation}
|\zeta>=\exp(\zeta a^{\dagger}b^{\dagger}-\zeta^{*}ab)|0,0>,
\end{equation}
In addition to these states many authours \cite{bar} \cite{gsa}  have 
considered the SU(1,1) coherent states of two mode systems or the
`pair coherent states' which were simultaneous eigenstates of $ab$ and 
$a^{\dagger}a-b^{\dagger}b$
\begin{eqnarray}
ab|\zeta,q>&=&\zeta|\zeta,q>,\nonumber \\
Q_{1}|\zeta,q>&=&q_{1}|\zeta,q>.
\end{eqnarray}
These can be mapped onto the SU(1,1) group by means of the two Boson 
realisation:
\begin{equation}
K_{1}^{+}=a^{\dagger}b^{\dagger} \;,\;
K_{1}^{-}=ab \; ,\;
K_{1}^{z}=\frac{1}{2}(a^{\dagger}a+b^{\dagger}b+1)\;,
\end{equation}
which form an SU(1,1) algebra with the commutation relations
\begin{equation}
\left[  K_{1}^{+},K_{1}^{-}\right]=-2K_{1}^{z} \;,
\left[  K_{1}^{z},K_{1}^{\pm}\right]=\pm K_{1}^{\pm}.
\end{equation}
The conservation law for $Q_{1}$ is related to the Casimir operator C for the 
SU(1,1) group;
which can be written as
\begin{equation}
C=\frac{1}{4}(1-(a^{\dagger}a-b^{\dagger}b)^{2})=\frac{1}{4}(1-Q_{1}^{2}).
\end{equation}
Thus the eigenstate of $Q_{1}$ is also an eigenstate of C and the pair 
coherent stateis related to the eigenstate of $K_{1}^{-}$ by Barut and 
Girardello.

These generate a representation $D^{q_{1}}$
 that correspond to the positive discrete series representation of 
 SU(1,1) \cite{bh}. In the number state basis, this corresponds to  the basis 
 states $|n_{1}+q_{1},n_{1}>$, where
\begin{equation}
|n_{1}+q_{1},n_{1}>=\frac{(a^{\dagger})^{n_{1}+q_{1}}(b^{\dagger})^{n_{1}}}
{((n_{1}!)(n_{1}+q_{1})!)^{\frac{1}{2}}}|0,0> ,
\end{equation}

The pair coherent state
in the number state basis labelled as $|\zeta_{1},q_{i}>$ is
\begin{equation}
|\zeta_{1},q_{1}>=N_{q_{1}}\sum_{n_{1}=0}^{\infty}\frac{\zeta_{1}^{n_{1}}}
{\sqrt{n_{1}!(n_{1}+q_{1})!}}|n_{1}+q_{1},n_{1}>,
\end{equation}
with
\begin{equation}
N_{q_{1}}=\left[ (|\zeta_{1}|)^{-q_{1}} I_{q_{1}}(2|\zeta_{1}|)\right] 
^{-1/2}\;.
\end{equation}
These states constitute a complete set in each sector $q_{i}$ and the 
completeness relation is given by
\begin{equation}
\int d^{2}\zeta_{1} \frac{2}{\pi} I_{q_{1}}(2|\zeta_{1}|)K_{q_{1}}(2|\zeta_{1}|)
|\zeta_{1},q_{1}><\zeta_{1}, q_{1}|
\end{equation}
for the normalized states .

We now consider the group obtained by the addition of two SU(1,1) generators 
defined for four modes a,b,c,d.
\begin{eqnarray}
K^{+}=a^{\dagger}b^{\dagger}+c^{\dagger}d^{\dagger}=K_{1}^{+}+K_{2}^{+}\;
,\nonumber \\
 K^{-}=ab+cd=K_{1}^{-}+K_{2}^{-}\;, \nonumber \\
 K^{z}=\frac{1}{2}(a^{\dagger}a+b^{\dagger}b+c^{\dagger}c+d^{\dagger}d+2)
 =K_{1}^{z}+K_{2}^{z}. \nonumber \\
C=\frac{(K^{+}K^{-} + K^{-}K^{+})}{2}-K_{z}^{2}.
\end{eqnarray}
The `bi- pair coherent states' or the coherent states for the Kronecker 
Product  are now the eigenstates of $K^{-}$ ,$C_{1}$, $C_{2}$ and $C$ .
If we restrict ourselves to the positive discrete series representations of 
SU(1,1)
then the Kronecker Product $D^{q_{1}} X D^{q_{2}}$
i.e the Clebsch Gordan series for SU(1,1) given by
\begin{equation}
D^{q_{1}}XD^{q_{2}}=\sum_{q=q_{1}+q_{2}+1}^{\infty} D^{q}.
\end{equation}
Thus a given representation in the Kronecker product is fixed by 
$q,q_{1},q_{2}$

The eigenvalue problem that we wish to solve is
\begin{equation}
K^{-}|\zeta,q>=\zeta|\zeta,q>\; \;\; ;
C|\zeta,q>=(1/4-q^{2}/4)|\zeta,q>.
\end{equation}
In terms of the product number state basis 
$|n_{1}+q_{1},n_{1}>|n_{2}+q_{2},n_{2}>$ we get:
\begin{eqnarray}
|\zeta,n,q_{1},q_{2}>&=&N_{n}\sum_{k=0}^{\infty}\frac{(\zeta)^{k}}
{\left[ (k)!(k+2n+q_{1}+q_{2}+1)!\right] ^{\frac{1}{2}}}\nonumber \\
&\times&\sum_{n_{1},n_{2}}C^{q_{1},q_{2},n}_{n_{1},n_{2},n+k}
\delta_{(n_{1}+n_{2},n+k)}|n_{1}+q_{1},n_{1}>|n_{2}+q_{2},n_{2}>\;.
\end{eqnarray}
 we get an expression for the Kronecker Product states in terms of the 
 CG coefficients in the photon number basis.
\subsection{Clebsh Gordan Problem in the pair coherent state basis}
Consider the four mode bases of the Hilbert space characterised by the 
product of two pair (SU(1,1) coherent states 
$|\zeta_{1},q_{1}> |\zeta_{2},q_{2}>$ .
Since these coherent states form an overcomplete set  any vector in the four 
mode Hilbert space can be expanded in terms of these states. In particular 
the coherent state of the product SU(1,1) X SU(1,1) $|\zeta, q>$ can be 
expanded directly in terms of
the unnormalized states
\begin{eqnarray}
|\zeta_{1},q_{1}>>&=&\sum_{n=0}^{\infty}\frac{\zeta_{1}^{n}}
{\sqrt{n!(n+q_{1})!}}|n+q_{1},n>\;, \nonumber \\
|\zeta_{2},q_{2}>>&=&\sum_{m=0}^{\infty}\frac{\zeta_{2}^{m}}
{\sqrt{m!(m+q_{2})!}}|m+q_{2},m>\;.
\end{eqnarray}
The completeness relation for the unnormalised  states $|\zeta_{i},q_{i}>>$ 
can be deduced from (2.18) to be
\begin{equation}
\int d^{2}\zeta_{i} \frac{2}{\pi} |\zeta_{i}|^{q_{i}}
K_{q_{i}}(2|\zeta_{i}|)|\zeta_{i},q_{i}>><<\zeta_{i},q_{i}|=1.
\end{equation}
The unnormalised states have the advantage that the operators $K_{i}^{\pm}$ and 
$K_{i}^{z}$
can be expressed as differential operators.
The completeness relation and resolution of the identity ensures that the 
product states $|\zeta_{1},q_{1}>>|\zeta_{2},q_{2}>>$ form the basis states 
for $D^{q_{1}}XD^{q_{2}}$ and any four mode state $|\psi>$ can be expanded as
\begin{equation}
|\psi>=\int <<\zeta_{1},q_{1}<<\zeta_{2},q_{2}|\psi> |\zeta_{1}q_{1}>> 
|\zeta_{2},q_{2}>> d^{2}\sigma(\zeta_{1}) d^{2}\sigma(\zeta_{2}).
\end{equation}
In this representation the quantity $<<\zeta_{1},q_{1},zeta_{2},q_{2}|\psi>$ 
is an analytic function $\psi(\zeta_{1}^{*},\zeta_{2}^{*},q_{1},q_{2})$ and 
the operators $K_{1}$ and $K_{2}$ act as ifferential operators on this 
function.
In particular the coherent state vector $|\zeta,q>$ in this four mode hilbert 
space can be written as:
\begin{equation}
|\zeta,q,q_{1},q_{2}>=\int <<\zeta_{1},q_{1}<<\zeta_{2},q_{2}|\zeta,q> 
|\zeta_{1}q_{1}>> |\zeta_{2},q_{2}>> d^{2}\sigma(\zeta_{1})d^{2}
\sigma(\zeta_{2}).
\end{equation}
This becomes the equivalent of the Clebsch Gordon equation in the pair 
coherent state basis and the quantity
The overlap function $<<\zeta_{1},q_{1}|\zeta_{2}q_{2}|\zeta,q>= 
f(\zeta_{1}^{*},\zeta^{*}_{2},\zeta q_{1},q_{2})$ is the equivalent of the 
Clebsch Godon coefiient for the SU(1,1) COHERENT STATE BASIS.
The action of the generators of SU(1,1) X SU(1,1) on f is given by
\begin{eqnarray}
(K_{1}^{+}+K_{2}^{+})f&=&(\zeta^{*}_{1}+\zeta^{*}_{2})f \nonumber \\
(K_{1}^{-}+K_{2}^{-})f&=&[(\frac{\partial}{\partial \zeta_{1}^{*}}
(q_{1}+\zeta_{1}^{*}\frac{\partial}{\partial \zeta_{1}^{*}} +
(\frac{\partial}{\partial \zeta_{2}^{*}}(q_{2}+\zeta_{2}^{*}
\frac{\partial}{\partial \zeta_{2}^{*}} ] f
\end{eqnarray}
On the other hand
\begin{equation}
K_f=\zeta f\;\;\; \; ;
Cf=[(1-q^{2})/4]f
\end{equation}
Thus we get the following two differential equations for f:

\[
\left[ \frac{\partial}{\partial\zeta_{1}^{*}}(q_{1}+\zeta_{1}^{*}\frac{\partial}
{\partial\zeta_{1}^{*}})
+\frac{\partial}{\partial\zeta_{2}^{*}}(q_{2}+\zeta_{2}^{*}\frac{\partial}
{\partial\zeta_{2}^{*}})\right]f =\zeta f\;,
\]
and
\begin{eqnarray}
\left[ \zeta_{1}^{*}\zeta_{2}^{*}(\frac{\partial^{2}}{\partial\zeta_{1}^{*2}}
-2\frac{\partial}{\partial \zeta_{1}^{*}}\frac{\partial}{\partial\zeta_{2}^{*}}
+\frac{\partial^{2}}{\partial\zeta_{2}^{2}})\right] f 
\nonumber \\
+\left[ (q_{1}+1)\zeta_{2}^{*}(\frac{\partial}{\partial \zeta_{1}^{*}}-
\frac{\partial}{\partial \zeta_{2}^{*}})
-(q_{2}+1)\zeta_{1}^{*}(\frac{\partial}{\partial \zeta_{1}^{*}}-
\frac{\partial}{\partial \zeta_{2}^{*}})\right] f   \nonumber \\
= -\left[ \frac{q^{2}}{4}-\frac{(q_{1}+ q_{2}+1)^{2}}{4}\right] f\;\nonumber
\end{eqnarray}

Solving these two equations we get \cite{ab}:
\begin{eqnarray}
f&=&<<\zeta_{1},q_{1},\zeta_{2},q_{2}|\zeta,q>>\nonumber \\
&=& N (\zeta(\zeta_{1}^{*}+\zeta_{2}^{*}))^{-q/2}I_{q}(\sqrt{4\zeta(\zeta_{1}^{*}
+\zeta_{2}^{*}}) (\zeta_{1}^{*}+\zeta_{2}^{*})^{n} P_{n}^{q_{2},q_{1}}
(\frac{\zeta_{1}^{*}-
\zeta_{2}^{*}}{\zeta_{1}^{*}+\zeta_{2}^{*}})
\end{eqnarray}
N is the normalisation .
Thus the state $|\zeta,q>$ can be obtained from the relation:
\begin{equation}
|\zeta,q>=\frac{4N}{\pi^{2}}\int d^{2}\zeta_{1}\int d^{2} \zeta_{2} 
K_{q_{1}}(2|\zeta_{1}|)K_{q_{2}}(2|\zeta_{2}|)<<\zeta_{1},q_{1},\zeta_{2},q_{2}|
\zeta,q>> |\zeta_{1},q_{1}>>|\zeta_{2},q_{2}>>
\end{equation}
This is the Clebsch Gordon form for the product basis of Coherent states of 
SU(1,1) X SU(1,1).

It is interesting to note that by
substituting the values of $|\zeta_{1},q_{1}>>$ and $|\zeta_{2},q_{2}>>$ 
given in equations (14)
and using the expansion for the Jacobi Polynomial as well as the expansion of the 
Bessel function $I_{q}$
and carrying out the various integrations we have:
\begin{eqnarray}
|\zeta,q>&=&N' \sum_{k=0}^{\infty} \frac{\zeta^{k}}{(k! (k+q)!)
^{\frac{1}{2}}} \sum_{n_{1},n_{2}} \delta_{(n_{1}+n_{2},n+k)}\nonumber\\
&&\left[ \frac{n_{1}! n_{2}!, (n_{1}+q_{1})! (n_{2}+q_{2})! k!}{(k+q)!}
\right]^{\frac{1}{2}} ((n+q_{1})!(n+q_{2})!)\nonumber \\
&& \sum_{l}(-1)^{l}\frac{1}{l!(q_{2}+l)!(n-l)!(n_{2}-l)!(n_{1}-n-l)!
(n+q_{1}-l)!} |n_{1}+q_{1},n_{1}>|n_{2}+q_{2},n_{2}>\;\;.
\end{eqnarray}
By comparison with expression [13] in the previous section we have:
\begin{eqnarray}
C_{n_{1},n_{2},n+k}^{q_{1},q_{2},n}&=&\left[ \frac{n_{1}! n_{2}!, 
(n_{1}+q_{1})! (n_{2}+q_{2})! k!}{(k+2n+q_{1}+q_{2}+1)!}\right]^{\frac{1}{2}} 
((n+q_{1})!(n+q_{2})!)^{1/2}\nonumber \\
&& \sum_{m}(-1)^{m}\frac{1}{(q_{2}+m)!(n-m)!(n_{2}-m)!(n_{1}-n-m)!(n+q_{1}-m)!}
\;\;.
\end{eqnarray}

Which is the Clebsch Gordon coefficient for the canonical number state basis 
for SU(1,1)XSU(1,1) .
\section{SubPoissonian Properties of SU(1,1)XSU(1,1) coherent states }
To give an idea of the Sub-Poissonian nature of these states let us consider
a special case which is useful in physical applications.
 Consider the case $q_{1}=q_{2}=0$ ; q=1 ; $\zeta\ne 0$\\
In this special case , we start with equal number of photons in the modes $a$ and $b$
and in $c$ and $d$.
Then
\begin{equation}
|\zeta, 1, 0, 0 >=N_{1}\sum_{k}\frac{ \zeta^{k}}{\left[(k+1)!(k)!\right] 
^{1/2}}\sum_{n_{1},n_{2}}
\frac{1}{(k+1)^{1/2}}\delta_{n_{1}+n_{2},k} |n_{1},n_{1}>|n_{2},n_{2}>\; ,
\end{equation}
where
\begin{equation}
N_{1}=\frac{(|\zeta|)^{1/2}}{\left[ I_{1}(2|\zeta|)\right]^{1/2}}\; \;  .
\end{equation}
 The single mode probability distribution $P_{n_{1}}$ and the mean number 
 of photons $<n_{1}>$ are given by
\begin{equation}
P_{n_{1}}(\zeta)=
 N_{1}^{2}|\zeta|^{2n_{1}}\sum_{n_{2}}\frac{|\zeta|^{2n_{2}}}
 {(n_{2}+n_{1}+1)!^{2}} 
 \; \; ,
\end{equation}
and
\begin{equation}
<n_{1}>=
 \frac{|\zeta|I_{2}(2|\zeta|)}{2I_{1}(2|\zeta|)}
\end{equation}
A measure of the non-classical nature of the distribution is given by Mandel's
Q parameter , which for the mode a is given by
\begin{eqnarray}
Q&=&\frac{<n_{1}^{2}>-(<n_{1}>)^{2}-<n_{1}>}{<n_{1}>}\\
&=&\frac{2 |\zeta|I_{3}(2 |\zeta|)}{3 I_{2}(2 |\zeta|)}-\frac{|\zeta| I_{2}(2 
|\zeta|)}{2 I_{1}(2|\zeta|)}.
\end{eqnarray}
In fig. 1 we plot Q .vs. $|\zeta|$.
For values of $|\zeta|<2$, Q is negative showing the departure from the 
Poisonnian.
The joint probability distribution $P_{n_{1}+n_{2}}$ can be calculated 
from $P_{n_{1},n_{2}}$ by the relation:
\begin{equation}
P_{k}=\sum_{n_{1},n_{2}}\delta_{n_{1}+n_{2},k}P_{n_{1},n_{2}}=\frac{N_{1}^{2} 
|\zeta|^{2k}}{k!(k+1)!}
\end{equation}

The average value $<k>$ is given by:
\begin{equation}
\sum_{k}kP_{k}=\frac{|\zeta|I_{2}(2|\zeta|)}{I_{1}(2|\zeta|)}
\end{equation}
In figure 2 we plot $P_{k}.vs. k$ and compare it to the corresponding Poissonian 
with mean value $<k>$ and it is clear that the distribution is sub Poissonian.
\section{Physical Applications}
 SU(1,1)XSU(1,1) states are useful states in dealing with physical systems 
 involving four modes of the radiation fields.
The physical problem could  be
the passage of two-beams of light each having  two  polarisation 
modes passing through a medium in which there is  a  competition 
between the non-linear gain due to an external pumping field and 
the non-linear absorption\cite{gsa86} \cite{zeil},\cite{schauer}.
 The states  generated
are precisely the states considered in this paper.
 Let each beam contain both left and  right
circularly polarised photons. Let a,b, a ,b  denote the creation 
and annilation operators for RIGHT circularly polarised  photons 
from beam 1 and beam 2 and $ c,d,c^{\dagger} ,d ^{\dagger}$denote  
the  creation  and
annihilation operators for LEFT circularly polarised photons  in 
beam 1 and beam 2.
The master equation describing the dynamic behaviour of the fields
resulting  from the competition between two photon  absorption 
and  four  wave mixing can be shown to be:
\begin{equation}
d\rho/dt=-K/2(O^{\dagger} O\rho -2O\rho O^{\dagger} 
+\rho O^{\dagger} O)-i\left[G(O^{\dagger} +O) , \rho\right]     [3]
\end{equation}
Where G denotes the four wave mixing susceptibility.
Where  K  is  related  to  the  cross-section  for  two   photon
absorption and O = ab+cd.
Defining an operator C=O+2iG/K
We have:
\begin{equation}
d\rho/dt=-K/2(C^{\dagger}C \rho+\rho C^{\dagger}C C-2C \rho C^{\dagger} )
\end{equation}
Whose steady state solution:
$C\rho=0$ with $\rho=|\psi><\psi|$ so that:
$C|\psi>=0$ implying that  $O|\psi>=-2iG/K|\psi>$
or $(ab+cd)|\psi>=\lambda|\psi>$
Where $\lambda=-2iG/K$
Thus the steady state  solutions  of  the  master  equations  are 
eigenstates of the operator O. Furthermore, if we now impose the 
condition that the initial state is one in which  the  difference 
in the in the the number of  photons  in  the  two  polarisation 
modes of each beam is a constant, with q  being the constant for 
the right circularly polarised photons and q being the constant 
for the left circularly polarised photon in beam 1 and beam 2,
the states $|\psi>$ are just the SU(1,1) X SU(1,1) coherent states.

Another
examples of processes where four modes of the radiation field are important 
involve
phase conjugate resonators  and  the process of
down conversion in the field of a standing pump wave\cite{boyd} .In the latter case,  
the forward wave will produce the
modes a and b and the backward pump will give the modes c and d. The 
Hamiltonian
for such interactions will have the form
\begin{equation}
H=(\epsilon_{f}^{*}ab+\epsilon_{b}^{*}cd + c.c),
\end{equation}
where $\epsilon_{f}$ and $\epsilon_{b}$ are the forward and backward
fields. Again the relevant coherent states are the eigenstates
of the operator
\begin{equation}
K^{-}=(ab+cd)=K_{1}^{-} + K_{2}^{-}.
\end{equation}
\section*{Acknowledgments}
I would like to thank the University Grants Commision, India, National Board 
of Higher Mathematics, India, Indian National Science Academy, 
DST and CSIR for financial support.

\end{document}